\documentclass[aps,prl,amsfonts,nofootinbib]{revtex4}

\usepackage{amssymb,amsmath}
\usepackage{graphicx}
\usepackage{epsfig}

\newcommand{\beq}{\begin{equation}}
\newcommand{\eeq}{\end{equation}}
\newcommand{\bea}{\begin{eqnarray}}
\newcommand{\eea}{\end{eqnarray}}
\newcommand{\vc}[1]{{\textbf{#1}}}
\newcommand{\mc}[1]{\mathcal{#1}}

\begin{document}

\hfill TUM 889/13
\vspace{0.5cm}

\title{Fluctuation-dissipation and equilibrium for scalar fields in de Sitter}

\author{Gerasimos Rigopoulos}
\affiliation{Physik Department T70, Technische Universit\"{a}t M\"{u}nchen, D--85748 Garching, Germany}

\begin{abstract}
\noindent The infrared dynamics of a minimally coupled scalar field in de Sitter spacetime can be described as Brownian motion of a particle in a medium of de Sitter temperature $T_{DS}=\frac{H}{2\pi}$. The system obeys a fluctuation-dissipation relation and its equilibrium distribution is Maxwell-Boltzmann, implying kinetic and potential energies of comparable magnitudes. The transition to equilibrium is a semi-classical process beyond the scope of perturbation theory for interacting fields. The stochastic kinetic energy of the field causes de Sitter spacetime to cool down slowly with a corresponding decrease of the effective vacuum energy.
\end{abstract}

\maketitle

\section{Introduction}
It has been known for a long time that spacetimes with horizons can be assigned a temperature and exhibit properties similar to thermodynamical systems \cite{Birrell:1982ix}. This connection has been extensively explored in the case of black holes which admit a description in terms of corresponding thermodynamical laws. Horizons also appear in cosmology. A prominent example is de Sitter space with horizon radius $R=1/H$ and temperature
\beq\label{DS-Temp}
T_{DS}=\frac{H}{2\pi}\,.
\eeq
One physical manifestation of this temperature is the thermal spectrum of excitations exhibited by an Unruh detector coupled to a scalar field in de Sitter, a phenomenon experienced by an observer confined within the de Sitter horizon. On the other hand, fluctuations of the scalar field itself on \emph{superhorizon} scales, highly relevant for inflation \cite{Mukhanov:2005sc}, do not exhibit obvious thermal properties. In fact, inflationary fluctuations have a scale invariant and not a thermal spectrum so there is no apparent link to thermodynamics and the temperature (\ref{DS-Temp}) in this case.

In this letter we establish a connection between scalar field fluctuations on superhorizon scales in de Sitter and thermodynamical considerations. We will be using the cosmologically relevant flat slicing coordinates in which the metric reads
\beq
ds^2=-dt+e^{2Ht}d\vc{x}^2\,,
\eeq
and assume that the the de Sitter phase starts at some definite time labeled by $t=0$ which will be assumed sufficiently earlier than all times considered here. The dynamics of a test scalar field is governed by the action
\bea\label{Model}
S=\int d^4x \, a^3 \Big[\frac{1}{2}\dot{\Phi}^2 - \frac{1}{2} \frac{\left(\partial_{i}\Phi\right)^2 }{a^2} - V(\Phi) \Big]\,,
\eea
leading to the classical Klein-Gordon equation
\beq\label{KG1}
\ddot{\phi} + 3H\dot{\phi}-e^{-2Ht}{\nabla^2}\phi+\frac{\partial V}{\partial\phi}=0\,,
\eeq
where $a=e^{Ht}$ is the scale factor. We will consider a light scalar for which the mass $m\ll H$.

On long wavelengths $k>aH$ the gradient term in (\ref{KG1}) becomes irrelevant and the motion of the scalar field is over-damped with the the expansion rate $H$ providing a friction term. Assuming an analogy can be drawn, let us recall some facts about known dissipative systems which exhibit friction on macroscopic scales. In such systems dissipation is generically accompanied by fluctuations with a firm relation between the fluctuation and the dissipation parameters. The earliest and most famous example is the phenomenon of Brownian motion described by the Langevin equation
\beq
m\frac{d\vc{v}}{dt}+\gamma\vc{v}=\vec{\xi}(t)
\eeq
where \vc{v} is the velocity of a Brownian particle with mass $m$ and $\gamma$ the friction coefficient. $\vec{\xi}(t)$ is a stochastic force due to the thermal motion of the medium in which the particle is immersed and whose correlator can be inferred from thermodynamic equilibrium
\beq
\langle\xi_i(t)\xi_j(t')\rangle=2\gamma T\delta(t-t')\delta_{ij}\,,
\eeq
where $T$ is the temperature of the medium and Boltzmann's constant is set to unity ($k_B=1$). Another famous example involves Johnson-Nyquist noise in electronic circuits. If $R$ is the resistance, $L$ the inductance and $C$ the capacitance of the circuit, the dynamics of charge $Q$ in the circuit is governed by the equation
\beq
L\frac{d^2Q}{dt^2} + R\frac{dQ}{dt}+\frac{Q}{C}=U_{\rm ext}+\xi_U(t)
\eeq
where $U_{\rm ext}$ is an externally applied voltage and $\xi_V(t)$ is an intrinsic omnipresent fluctuating voltage related to thermodynamical fluctuations of the electromagnetic fields. Statistical physics again predicts that
\beq
\langle \xi_U(t)\xi_U(t')\rangle=2RT\delta(t-t')
\eeq
In both cases the magnitude of the friction term in the dynamical equations (dissipation) determines along with the temperature the amplitude of the thermally induced random force (fluctuation). This is a facet of the (classical) fluctuation-dissipation theorem of statistical physics.

Given the omnipresence of fluctuation-dissipation relations in nature, one can conjecture by analogy that a fluctuation term should necessarily accompany the scalar field dynamics (\ref{KG1}) on long wavelengths where the friction dominates. We thus expect that the true dynamics of the scalar field on physical scales $r>1/H$ ($r=e^{Ht}x$) to be governed by
\beq\label{KG2}
\ddot{\phi}+3H\dot{\phi}+\frac{\partial V}{\partial\phi}=\xi(t)
\eeq
where the amplitude of the fluctuation term should conform to a fluctuation dissipation relation
\beq\label{KG-noise}
\langle\xi(t)\xi(t')\rangle = \frac{(2\times 3H \times T_{DS})}{\frac{4\pi}{3}\left(\frac{1}{H}\right)^3}\,\delta(t-t')\,.
\eeq
The numerator directly mirrors the known fluctuation-dissipation relations described above. The inverse volume factor is required for dimensional reasons; we will find below that it is simply the physical volume of the de Sitter horizon $R_{\rm DS}=1/H$;

The above equation shows that the ``soft'' part of the field $\phi$, comprised of modes with physical wavelengths $k/a<1/R_{\rm DS}$, executes Brownian motion in an environment of temperature $T_{\rm DS}$. It should be stressed that no external thermal environment is assumed and $\phi$ is in its ground state, taken to be the Bunch-Davis vacuum. The source of the fluctuations are the \emph{vacuum} fluctuations of $\phi$ itself, which are amplified on long scales, and the thermal behavior of the soft modes arises solely from the effects of spacetime curvature on the quantum field. In what follows we verify (\ref{KG2}) and (\ref{KG-noise}) as the correct long wavelength description by explicitly integrating out the ``hard'' modes of the field with physical wavelengths shorter than $R_{\rm DS}$.

The idea that the dynamics of long wavelength light fields in (quasi-)de Sitter can be described by a stochastic equation was first explicitly proposed by Starobinsky \cite{Starobinsky:1986fx} (see also \cite{Starobinsky:1994bd}). The treatment presented here, based on out-of-equilibrium methods of statistical physics, clarifies and extends Starobinsky's original stochastic inflation approach. The agreement of the stochastic approach with Quantum Field Theory at the perturbation level was first pointed out in \cite{Tsamis:2005hd}. Here, we exhibit this link in full generality and at the non-perturbative level for a scalar field. We show that the thermodynamic analogy alluded to above is exact and the equilibrium distribution of the scalar field is Maxwell-Boltzmann. The latter implies an average kinetic energy comparable to the average potential energy with equipartition for free fields. Such a relation is characteristic of thermodynamic equilibrium no matter how strong the dissipation. These results allow for an exact thermodynamic interpretation of the de Sitter temperature for the case of cosmologically relevant scalar field fluctuations.

\section{Stochastic dynamics in DeSitter}
To prove the assertions made above we proceed to obtain an effective theory for the long wavelength modes of the scalar field. We split $\phi$ into a long wavelength part $\phi(\vc{k} \lesssim \epsilon aH)$ and a short wavelength part $\varphi(\vc{k} \gtrsim \epsilon a H)$: $\Phi=\phi+\varphi$, where $\epsilon$ is a small parameter that controls the separation between the long and short sectors. As long as $\epsilon\ll 1$ our final result will be independent of $\epsilon$.

The split is achieved by a window function $W_k(t)$ in k-space to be discussed shortly. The contributions to the action from $\phi$ and $\varphi$ are
\beq
S=S[\phi]+\int d^4x\, a^3\left[\frac{1}{2}\,\varphi \hat{\mathcal{O}}\,\varphi
+\varphi\left(\hat{\mathcal{O}}\phi\right)\right]+ S_{\rm int}[\varphi,\phi]
\eeq
where $\hat{\mathcal{O}}=\left(\partial_t^2+3H\partial_t-\frac{\nabla^2}{a^2}+m^2\right)$.
$S[\phi]$ is the complete action for the long wavelength fields, \emph{including all non-linear terms involving the long field $\phi$}. We have collected all other interaction terms in $S_{\rm int}$. The second term is the free action for the short field and a bilinear long-short coupling controlled by $\ddot{W}_k(t)+3H\dot{W}_k(t)$; bilinear coupling terms that do not involve time derivatives of the window function vanish. This coupling arises solely from the time dependence of the long-short split and describes the influence of short wavelength modes that cross into the long wavelength sector due to the cosmic expansion. The form of this coupling term suggests that we define the long wavelength sector by the window function
\beq\label{window}
W_k(t)=\left(1-\frac{k^3}{(\epsilon a H)^3}\right)\frac{1}{H}\,\Theta\!\!\left[\ln\left(\frac{\epsilon a H}{k}\right)\right]
\eeq
for which $\ddot{W}_k(t)+3H\dot{W}_k(t)=\delta(t-\frac{1}{H}\ln\left(k/\epsilon H\right))$. Thus, the coupling is effective for any mode $k$ when $k=\epsilon a H$.

The effective long wavelength action $A[\phi]$ is defined by integrating out the short field $\varphi$ in the Closed-Time-Path (CTP) formalism
\beq\label{A-with-int}
e^{i A_\mc{C}[\phi] }= e^{iS_\mathcal{C}[\phi]}\int
D\varphi_{\mc{C}}\,e^{i\int_\mc{C}\mathcal{L}_{\rm int}\left[\phi,\varphi\right]} \,e^{i\int_{\mathcal{C}}\, a^3\left[\frac{1}{2}\varphi \hat{\mathcal{O}}\varphi
+\varphi\left(\hat{\mathcal{O}}\phi\right)\right]}
\eeq
where the subscript $\mc{C}$ indicates integration along the closed Keldysh (forward-backward) time contour - see \cite{Morikawa:1989xz} for an early treatment at leading order. The $\varphi$ integration can be performed with the term containing $\mathcal{L}_{\rm int}$ handled perturbatively. Its treatment will be presented elsewhere. Ignoring $\mathcal{L}_{\rm int}$ for now we obtain
\bea\label{MSRJD-1}
A[\phi,\psi]=&&\int d^4\tilde{x} \,a^3\left[-{\psi}\left(\ddot{\phi} +3H\dot\phi-{\frac{H^2\epsilon^2}{a^2}{\nabla}_{\tilde{\vc{x}}}^2\phi}+\frac{\partial V}{\partial\phi}\right) +  2\!\sum\limits_{m=1}^\infty\!\frac{V^{{(2m+1)}}}{\left(2m+1\right)!}\left(\frac{{\psi}}{2}\right)^{2m+1}\!\left( \epsilon H\right)^{6m}\right]\nonumber\\
&&+\frac{i}{2}\frac{9H^5}{4\pi^2}\int d^4\tilde{x} d^4\tilde{x}' \, a^3(t)a^3(t')\,{\psi}(\tilde{x})\mc{N}(\tilde{x},\tilde{x}'){\psi}(\tilde{x}')
\eea
where $\tilde{\vc{x}}=\vc{x}\epsilon H$ and we defined $\phi=\frac{1}{2}\left(\phi_++\phi_-\right)$, $\psi=\frac{\left(\phi_+-\phi_-\right)}{\left(\epsilon H\right)^3}$, $\mc{N}(\tilde{x},\tilde{x}')=\frac{\sin\left(a \left|\tilde{\vc{x}}-\tilde{\vc{x}}'\right|\right)}{a \left|\tilde{\vc{x}}-\tilde{\vc{x}}'\right|}\delta(t-t')$. For $\epsilon\ll 1$, ie considering wavelengths much larger than $1/H$, the relevant terms are those linear and quadratic in the field $\psi$ without the gradient term. Approximating $\mc{N}(x,x')\rightarrow \delta(t-t')\frac{\delta(\tilde{\vc{x}}-\tilde{\vc{x}})}{a^3}$ we therefore find that long wavelength correlation functions can be computed using\footnote{This is known as an MSRJD functional after Martin, Siggia, Rose, Janssen and deDominicis \cite{Martin:1973zz, Altland:2006si}}
\beq\label{MSRJD-2}
\langle\mc{Q}(\phi)\rangle=\prod\limits_{\tilde{\vc{x}}}\int [D\phi] [D\psi] \,\, \mc{Q}(\phi)\,\,
e^{i\int dt a^3\left[-\psi\left(\ddot{\phi} +3H\dot\phi+\frac{\partial V}{\partial\phi}\right) +\frac{i}{2}\frac{9H^5}{4\pi^2} \psi^2\right]}\,.
\eeq
Different spatial points are uncorrelated and in what follows we will be focusing on the functional for a single point. It should of course be remembered that in reality each ``point'' corresponds to a region of physical size $R\sim 1/H$.

The functional integral (\ref{MSRJD-2}) can be used for perturbative calculations. Writing
\beq\label{MSRJD-pert}
A_{\tilde{\vc{x}}}=\frac{1}{2}\int dt\, \left[ \left(\begin{smallmatrix}\phi\,, &\psi \end{smallmatrix}\right)\left(\begin{smallmatrix}0 & -a^3(\partial_t^2+3H\partial_t+m^2)\\ -a^3(\partial_t^2+3H\partial_t + m^2) & ia^3\,\frac{9H^5}{4\pi^2}\end{smallmatrix}\right)
\left(\begin{smallmatrix}\phi \\ \psi \end{smallmatrix}\right) + a^3\lambda\psi\phi^3\right]
\eeq
we  obtain the free correlation functions
\beq\label{propagators}
\left(\begin{smallmatrix}\langle\phi(t)\phi(t')\rangle & \langle\phi(t)\psi(t')\rangle \\ \langle\psi(t)\phi(t')\rangle & \langle\psi(t)\psi(t')\rangle\end{smallmatrix}\right)=
\left(\begin{smallmatrix}
F(t,t')&-iG^R(t,t')\\
-iG^A(t,t')&0\end{smallmatrix}\right)
\eeq
where $G^{(R,A)}(t,t')$ are the retarded and advanced Green functions
\beq
G^R(t,t')=G^A(t',t)=\frac{1}{3H}\frac{1}{a^3(t')}
\left(e^{-\frac{m^2}{3H}(t-t')}-e^{-3H(t-t')}\right)\Theta(t-t')
\eeq
and
\beq\label{corr}
F(t,t')=\frac{9H^5}{4\pi^2}\int\limits_{0}^{+\infty} d\tau \,a^6(\tau)\,\,G^R(t,\tau)G^A(\tau,t')
\eeq
Thus
\beq
\langle\phi(t)\phi(t')\rangle\simeq \frac{3H^4}{8\pi^2m^2}e^{-\frac{m^2}{3H}|t-t'|}\,.
\eeq
In the massless limit $G^R(t,t')\rightarrow \frac{1}{3H a^3(t')}\left(1-e^{-3H(t-t')}\right)\Theta(t-t')$
and the variance grows linearly with time
\beq
\langle\phi^2(t)\rangle_{m=0}\simeq \frac{H^3}{4\pi^2}t.
\eeq
It is  easy to obtain the higher order loop corrections at the coincident limit. From the propagators in (\ref{propagators}) and the vertex in (\ref{MSRJD-pert}) we have at one loop
\beq
\langle\phi(t)^2\rangle=F(t,t)+i(-i)\lambda\int d\tau \, a^3(\tau)\Big(F(t,\tau)+F(\tau,t)\Big)G^R(t,\tau)F(\tau,\tau)\,,
\eeq
which implies that in the massless case perturbation theory breaks down when $Ht \sim \frac{1}{\sqrt{\lambda}}$, while in the massive case this occurs at $Ht\sim \frac{m^2}{\lambda H^2}$. From these results it is clear that perturbation theory is inadequate for describing the evolution of the interacting system for arbitrarily large times and in particular the eventual transition to equilibrium which we describe below.

\section{The equilibrium distribution}
The long wavelength dynamics represented by (\ref{MSRJD-2}) corresponds to a stochastic evolution of the field $\phi$. This is made clear by performing a Hubbard-Stratonovich transformation
\beq
e^{-\int dt\,a^3\frac{9H^5}{8\pi^2} {\psi}^2 }=\int[D\xi]e^{-\int dt\,a^3\left[\frac{2\pi^2}{9H^5}\xi^2-i\xi{\psi}\right]}
\eeq
leading to
\beq
\langle\mc{Q}(\phi)\rangle=\int[D\xi]e^{-\int dt\,a^3\left[\frac{2\pi^2}{9H^5}\xi^2\right]}
\int[D\phi]\mc{Q}(\phi)\,\delta\left(\ddot{\phi}+3H\dot{\phi}+\frac{\partial V}{\partial\phi}-\xi\right)
\eeq
Thus, the field at each spatial point satisfies a Langevin equation
\beq\label{Langevin}
\ddot{\phi}+3H\dot{\phi}+\frac{\partial V}{\partial\phi}=\xi
\eeq
where the noise term is gaussian with
\beq\label{noise-2}
\langle\xi(t)\xi(t')\rangle=\int[D\xi]\xi(t)\xi(t')e^{-\int dt\,a^3\left[\frac{2\pi^2}{9H^5}\xi^2\right]}=\frac{9H^5}{4\pi^2}\delta(t-t')\,.
\eeq
This shows that expectation values can be obtained by averaging over different stochastic histories of the field. It is also evident that (\ref{Langevin}) and (\ref{noise-2}) coincide with the assertions made in the introduction. Furthermore, the Langevin dynamics suggests that an equilibrium should be eventually reached if the potential is bounded from below as was first suggested in \cite{Starobinsky:1994bd}.

To obtain the probability distribution for $\phi$ we consider the quantity
\bea
\mc{P}\left(\phi,t|\phi_i,t_i\right)&=&\int\limits_{\phi_i}^{\phi_f} [D\phi] [D\psi] \,\,
e^{i\int^t_{t_i} dt\left[-\psi\left(\ddot{\phi} +3H\dot\phi+\frac{\partial V}{\partial\phi}\right)+\frac{i}{2}\frac{9H^5}{4\pi^2} \psi^2 \right]}\\
&=&\int\limits_{\phi_i}^{\phi_f} [D\phi][Dy][D\psi][D\rho] \,\,
e^{\int^t_{t_i} dt\left[-i\rho\left(\dot{\phi}-y\right)-i\psi\left(\dot{y} +3Hy+\frac{\partial V}{\partial\phi}\right)-\frac{1}{2}\frac{9H^5}{4\pi^2} \psi^2 \right]}
\eea
which represents the \emph{probability}\footnote{This is inherited from the closed time path contour in the initial path integral.} to find the field value $\phi$ at time $t$, given the field value $\phi_i$ at $t_{\rm i}$ in any such cell. Writing $\psi=-ip$, $\rho=-iq$
we have
\beq
\mc{P}\left(\phi,t|\phi_i,t_i\right)=\int\limits_{\phi_i}^{\phi_f} [D\phi][Dy][Dp][D\rho] \,\,
e^{-\int^t_{t_i} dt\left[q\dot{\phi}+p\dot{y}-H(p,q,y,\phi) \right]}
\eeq
where the ``Hamiltonian'' is
\beq
H(p,q,y,\phi)=\frac{9H^5}{8\pi^2}p^2-p\left(3Hy+\frac{\partial V}{\partial\phi}\right)+qy
\eeq
With $p=-\partial_y$ and $q=-\partial_\phi$ and normal ordering in the Hamiltonian, the probability $\mc{P}$ will satisfy a corresponding ``Shr\"{o}dinger'' equation
which is nothing but the Fokker-Planck equation
\beq\label{FP1}
\partial_t\mathcal{P}=\left(\frac{9H^5}{8\pi^2}\frac{\partial^2}{\partial y^2}+3H\frac{\partial}{\partial y}y+\frac{\partial V}{\partial\phi}\frac{\partial}{\partial y}-y\frac{\partial}{\partial\phi}\right)\mc{P}
\eeq
The correlation functions generated by (\ref{MSRJD-2}) can thus be computed using $\mc{P}$. Equilibrium is described by the stationary solution to (\ref{FP1}) which is easily found to be
\beq\label{MB-dist}
\mc{P}(\phi,y)=\frac{e^{-\frac{8\pi^2}{3H^4}\frac{y^2}{2}}e^{-\frac{8\pi^2}{3H^4}V}}{N}
\eeq
with
\beq
N=\frac{2\sqrt{\pi}}{\sqrt{3}H^2}\int d\phi \,\,e^{-\frac{8\pi^2}{3H^4}V}
\eeq
Therefore, any late time correlation function $\langle\mc{O}(\phi,y)\rangle$ \emph{in equilibrium} can then be written
\beq\langle\mc{O}(\phi,y)\rangle
=\int d\phi d{y} \,\,\mc{O}(\phi,y)\,\,\frac{e^{-\frac{8\pi^2}{3H^4}(\frac{y^2}{2}+V)}}{N}
\eeq
where $y=\dot{\phi}$. The system therefore equilibrates to a Maxwell-Boltzmann distribution.

\section{Discussion}

The long wavelength part of a scalar field in de Sitter, averaged over a region of physical radius $R\sim\frac{1}{H}$, performs Brownian motion in a medium of temperature $T_{DS}=\frac{H}{2\pi}$. A fluctuation-dissipation relation holds between the amplitude of the fluctuations and the ``friction'' term in the Klein Gordon equation. An equilibrium is reached, given by (\ref{MB-dist}) which is nothing but the Maxwell-Boltzmann distribution for the kinetic and potential energy in a volume $\frac{4\pi}{3}R_{DS}^3$ at temperature $T_{DS}$. No thermal environment is a priori assumed and the Brownian motion emerges naturally from the effects of spacetime curvature on the quantum field. It is interesting to note that the long wavelength dynamics is semiclassical in that the terms suppressed by $\epsilon$ in (\ref{MSRJD-1}) are also suppressed by powers of $\hbar^{2m}$, as can be seen by reinstating $\hbar$ and rescaling $\psi\rightarrow\hbar\psi$ \cite{Altland:2006si}. However, quantum physics affects long wavelengths through the $\psi^2$ fluctuation term in (\ref{MSRJD-2}) which induces stochastic noise.

Although these results are in agreement with earlier stochastic treatments on the part of the distribution $\propto e^{-\frac{8\pi^2}{3H^4}V}$, they also predict a significant average kinetic energy for the field. Equipartition is predicted for a free field, as can also be seen from (\ref{corr}), while $V=\frac{\lambda}{4}\phi^4$ gives
\beq\label{energy}
\frac{1}{2}\langle\dot{\phi}^2\rangle=2\langle V\rangle= \frac{3H^4}{16\pi^2}\,.
\eeq
The magnitude of the kinetic energy (\ref{energy}) does not of course contradict the fact that field modes are frozen when they enter the long wavelength sector. This kinetic energy reflects the change induced in the long wavelength field from the continuous influx of horizon crossing modes and is a result of choosing a fixed physical and not comoving UV cutoff for the long wavelength theory. It does imply however that the slow-roll assumption cannot be made for the long wavelength field subject to fluctuations in this case. This is a generic result of dissipative systems: no matter how strong the dissipation, fluctuations work to restore equipartition between potential and kinetic energy.

A few comments are in order regarding the above results. Firstly, non-perturbative calculations in \emph{Euclidean} de Sitter for $V=\frac{\lambda}{4}\phi^4$ \cite{Rajaraman:2010xd, Beneke:2012kn} agree with the equal time expectation values of operators $\mc{O}(\phi)$ obtained using (\ref{MB-dist}). The agreement goes beyond the leading order in powers of $\sqrt{\lambda}$ if (\ref{A-with-int}) is expanded. This will be discussed in more detail in forthcoming work. Thus, it would seem that a scalar test field in Lorentzian de Sitter can be thought of as a non-equilibrium thermodynamic system that relaxes towards equilibrium which in turn can also be described in Euclidean de Sitter. A second comment relates to the infrared divergences that plague cosmological computations \cite{Seery:2010kh}. We saw that in the present context such divergences are simply artifacts of perturbation theory and are removed if the semi-classical non-perturbative nature of the infrared sector is taken into account. Different approaches also suggest that such infrared divergences can indeed tamed in de Sitter and physically interpreted \cite{Riotto:2008mv, Burgess:2009bs, Garbrecht:2011gu, Boyanovsky:2012nd, Serreau:2013psa,Lazzari:2013boa}. One would expect that similar findings would hold for more realistic quasi-de Sitter inflationary models, also including scalar gravitational perturbations. These issues will be explored elsewhere.

Before closing we would like to note that all of the above results completely ignore the backreaction of the scalar field on the spacetime. However, if the average energy momentum tensor obtained from (\ref{energy}) is naively plugged into the Friedman equations one obtains that the rate of expansion slows down according to
\beq
H(t)=\frac{H}{\left(\frac{9}{8\pi^2}\frac{H^3}{M_p^2}t+1\right)^{\frac{1}{3}}}\,,
\eeq
leading to a decrease of the expansion rate and the temperature. Since the rate of change is small it is reasonable to assume that the stochastic analysis holds up even in this case. Thus, the induced stochastic kinetic energy of the field causes de Sitter spacetime to cool down slowly and the effective vacuum energy to decrease. For a massive particle this can continue only up to the point where the temperature becomes comparable to its mass after $t\sim \frac{8\pi^2}{9}\frac{M_p^2}{m^3}$.

\section{Acknowledgements}
\noindent The author would like to thank Martin Beneke and Bjorn Garbrecht for enlightening conversations. This work is supported by the Gottfried Wilhelm Leibniz programme of the Deutsche Forschungsgemeinschaft (DFG)


\begin{thebibliography}{99}

\bibitem{Birrell:1982ix}
  N.~D.~Birrell and P.~C.~W.~Davies,
  Cambridge, Uk: Univ. Pr. ( 1982) 340p

\bibitem{Mukhanov:2005sc}
  V.~Mukhanov,
  Cambridge, UK: Univ. Pr. (2005) 421 p

\bibitem{Starobinsky:1986fx}
  A.~A.~Starobinsky,
  In *De Vega, H.j. ( Ed.), Sanchez, N. ( Ed.): Field Theory, Quantum Gravity and Strings*, 107-126

\bibitem{Starobinsky:1994bd}
  A.~A.~Starobinsky and J.~Yokoyama,
  Phys.\ Rev.\ D {\bf 50} (1994) 6357
  [astro-ph/9407016].

\bibitem{Tsamis:2005hd}
  N.~C.~Tsamis and R.~P.~Woodard,
  Nucl.\ Phys.\ B {\bf 724} (2005) 295
  [gr-qc/0505115].

\bibitem{Morikawa:1989xz}
  M.~Morikawa,
  Phys.\ Rev.\ D {\bf 42} (1990) 1027.

\bibitem{Martin:1973zz}
  P.~C.~Martin, E.~D.~Siggia and H.~A.~Rose,
  Phys.\ Rev.\ A {\bf 8} (1973) 423.

\bibitem{Altland:2006si}
  A.~Altland and B.~Simons,
  Cambridge, UK: Univ. Pr. (2010), 2nd ed, 786 p

\bibitem{Rajaraman:2010xd}
  A.~Rajaraman,
  Phys.\ Rev.\ D {\bf 82} (2010) 123522
  [arXiv:1008.1271 [hep-th]].

\bibitem{Beneke:2012kn}
  M.~Beneke and P.~Moch,
  arXiv:1212.3058 [hep-th].

\bibitem{Seery:2010kh}
  D.~Seery,
  Class.\ Quant.\ Grav.\  {\bf 27} (2010) 124005
  [arXiv:1005.1649 [astro-ph.CO]].

\bibitem{Riotto:2008mv}
  A.~Riotto and M.~S.~Sloth,
  JCAP {\bf 0804} (2008) 030
  [arXiv:0801.1845 [hep-ph]].

\bibitem{Burgess:2009bs}
  C.~P.~Burgess, L.~Leblond, R.~Holman and S.~Shandera,
  JCAP {\bf 1003} (2010) 033
  [arXiv:0912.1608 [hep-th]].

\bibitem{Garbrecht:2011gu}
  B.~Garbrecht and G.~Rigopoulos,
  Phys.\ Rev.\ D {\bf 84} (2011) 063516
  [arXiv:1105.0418 [hep-th]].

\bibitem{Boyanovsky:2012nd}
  D.~Boyanovsky,
  Phys.\ Rev.\ D {\bf 86} (2012) 023509
  [arXiv:1205.3761 [astro-ph.CO]].

\bibitem{Serreau:2013psa}
  J.~Serreau and R.~Parentani,
  arXiv:1302.3262 [hep-th].

\bibitem{Lazzari:2013boa}
  G.~Lazzari and T.~Prokopec,
  arXiv:1304.0404 [hep-th].


\end{thebibliography}
\end{document}